# PRE-render Content Using Tiles (PRECUT). 1. Large-Scale Compound-Target Relationship Analyses

Sung Jin Cho*

*CimplSoft, Thousand Oaks, CA  91320.*

*Corresponding author phone: (805) 807-3913; e-mail: sjcho@cimplsoft.com.

**ABSTRACT**

Visualizing a complex network is computationally intensive process and depends heavily on the number of components in the network. One way to solve this problem is not to render the network in real time. **PRE**-render **C**ontent **U**sing **T**iles (**PRECUT**) is a process to convert any complex network into a pre-rendered network. Tiles are generated from pre-rendered images at different zoom levels, and navigating the network simply becomes delivering relevant tiles. PRECUT is exemplified by performing large-scale compound-target relationship analyses. Matched molecular pair (MMP) networks were created using compounds and the target class description found in the ChEMBL database. To visualize MMP networks, the MMP network viewer has been implemented in COMBINE and as a web application, hosted at http://cheminformatic.com/mmpnet/.

**INTRODUCTION**

Complex systems are everywhere. We study and exploit them and often end up with further complicating an already complex system. In drug discovery, drug-drug, drug-protein, and protein-protein interactions are components found in one such system. Drug discovery researchers study various interactions to identify abnormal ones and investigate a way to remove or counteract them to restore the system to a normal state. Perturbing the system using different chemical entities to observe their effects is one way to understand drug-protein interactions and leads to synthesizing a large number of small molecules.[1] Unfortunately, efforts to understand often lead to adding more components as in this case.

Components and relationships between them in a complex system are described as a network consists of nodes (objects) and edges (relationships). Many commercial and open source tools are available to study and visualize complex networks,[2,3,4,5] and with the advent of big data, more will be needed to uncover hidden relationships.  One common problem most of these tools are facing is that their performances are heavily depend on the number of components in the system.  One way to remove this dependency is not to render them in real time.  Examples of such visualization systems that can handle a large amount of data by pre-rendering display contents are various browser-enabled mapping applications like Google Maps.[6,7,8]  Contents to be displayed, maps in these applications, are pre-rendered at different zoom levels, and tiles are generated from the pre-rendered images to deliver requested information efficiently.  By only providing tiles of interest and overlaying relevant information, users can navigate to any part of the map quickly and retrieve requested information fast.  Any content that can be pre-rendered can be visualized this way including chemical and biological networks.

In this paper, **PRE**-render **C**ontent **U**sing **T**iles (**PRECUT**), a process to convert any complex network into a pre-rendered network, is described.  To illustrate PRECUT, matched molecular pair (MMP) networks were created to perform large-scale compound-target relationship analyses.  To visualize MMP networks, the MMP network viewer has been implemented in COMBINE[9] and as a web application, hosted at http://cheminformatic.com/mmpnet/.

**RESULTS and DISCUSSIONS**

**PRECUT**

A flow chart describing the PRECUT process is shown in Figure 1, and how this process is applied to generate a MMP network is illustrated in Figure 2.  A dataset is first processed, and relevant information are extracted to form pairs.  A pair consists of any two linked objects, and both objects and their linkage are defined by users.  Examples of the pair are compound-substructure hierarchy, compound-compound similarity, protein-protein interaction, etc.  A layout algorithm is used to generate graphs and the 2D representation of them. The 2D representation at each zoom level is converted to an image which is used to generate tiles. A tile viewer is used to visualize tiles and overlaid information. The dashed arrows in Figure 1 represent the data that are supplied to the tile viewer. To generate MMP networks, ChEMBL compounds and the target class description found in the ChEMBL database were used (Figure 2a-b). The large graph layout (LGL)[10] program[11] was used to construct undirected graphs and 2D coordinates of nodes (Figure 2c). Tiles were generated from images rendered using the LGL output (Figure 2d-e).

**Dataset**

After Babel[12] was used to convert compounds in the ChEMBL database from SMILES to MOL format, those that do not violate "the rule of 5" were selected and assigned their intended target classes. An intended target class is assigned when a compound is tested against an assay that are linked to the target class. The intended target classes are used to describe intentions behind syntheses and the assay selection. The intended target class assignment is used to let users know that an activity data exists, and further exploration is needed to assess the nature of the activity.

**Matched Molecular Pair**

A matched molecular pair is defined as a pair of molecules with a small structure difference.[13] It is a useful cheminformatics technique to examine how changes between two molecules can affect their activities. MMPs were generated using the method reported by Hussain and Rea.[14] Briefly, fragments are generated by, first, identifying all acyclic single bonds formed between two non-hydrogen atoms. The second step is to enumerate all possible single cuts, double cuts, and triple cuts to generate fragments (Figure 3). Finally, fragments are indexed by systematically creating key-value pairs. The "key" fragments are ones found in both compounds. The "value" fragments are ones that are different.

The number of non-hydrogen atoms found in the "value" fragment (set to 10) is used to control the biggest change allowed. 14,680,477 MMs were generated from 782,524 ChEMBL compounds (Figure 2a-b). Both fragment generating and indexing routines were written using c/c++ and incorporated into COMBINE.[9] RDKit[15] (for fragment generating) and Perl (for indexing) were also used to create the web version.

**Large Graph Layout**

The LGL program[11] uses a force-directed graph drawing algorithm to generate a layout. Attractive forces were applied to pull connected nodes toward each other, and repulsive forces were applied to push away proximal nodes. To deal with a large network, the LGL program first identifies connected sets that make up the original network. After layouts are generated for each connected set, the program merges them into one coordinate system, putting larger ones in the center and spreading smaller ones outward. The LGL program also reduces the number of edges used to calculate attractive forces by only including edges that form the minimum spanning tree (MST)[16] of the network. The MST edges were used to create network images to reduce visual clutter.

**Image Processing**

Seven different pixel images, 256x256, 512x512, 1024x1024, 2048x2048, 4056x4056, 8192x8192, and 16384x16384, were generated for zoom levels 0, 1, 2, 3, 4, 5, and 6, respectively. 256x256 pixel tiles were cut out of each image. This cut out process produces 1, 4, 16, 64, 256, 1024, and 4096 tiles. Figure 2d shows 256x256, 512x512, and 1024x1024 pixel images representing zoom levels 0, 1, and 2, respectively, and Figure 2e shows 1, 4, and 16 tiles generated from them. How generated tiles are used is illustrated in Figure 3.

**MMP Network Viewer**

Generated tiles can be viewed using the MMP network viewer hosted at http://cheminformatic.com/mmpnet/ or using COMBINE[9] (Figures 2f-g and 3). The first MMP network was generated using 782,524 compounds with 14,680,477 MMPs (All network). To enlarge the connected set at the center of the previous MMP network, smaller sets were removed, and images and tiles were regenerated (Center network). The center network contains 341,052 compounds with 12,050,339 MMPs. GPCR and kinase target classes are the two biggest classes, and GPCR and kinase intended compounds were used to generate GPCR and Kinase networks containing 253,024 compounds with 5,041,391 MMPs and 61,405 compounds with 567,488 MMPs, respectively. Finally, ring fragments generated after applying single, double, and triple cuts were filtered to remove infrequent ring fragments (found in less than 10 compounds). Compounds with multiple intended target classes were also removed. The resulting network contains 260,711 compounds with 2,605,526 MMPs (Ring network).

Figure 5 shows how chemical and biological data can be added to each map using the web application. This is illustrated by identifying a MMP interactively using CHEMBL446019 as a starting point to find its MMP counterpart, CHEMBL49986. Double clicking the node representing the first compound in the MMP network (or using the Find button) adds the structure of CHEMBL446019. A line forms between the node and the structure. When the MMP network is selected, users can move both the structure and the MMP network together using the left mouse button. When the structure is selected, it is moved independently respect to the MMP network. Double clicking the structure opens a web page listing known activities and references. Pressing the button displaying the structure generates fragments, and ChEMBL compounds containing a selected fragment are returned. Since a MMP is formed by identifying shared fragments, any shared fragment can lead to a MMP. There are two links associated with each structure. The top one opens the compound information page found in the ChEMBL web site. The bottom one opens the MMP network viewer, and the structure of the selected compound is displayed next to the node representing it. The automatic fragmentation process enhances the user experience because users can focus on the task at hand and do not have to spend time trying to draw a substructure. Unfortunately, each double-click or click action opens a new webpage, and the number of web pages created increases quickly. The web application could be designed to replace the current web page rather than open a new one to reduce the number of web pages, but it was decided that having the previous web page is more important.

The same task can be performed using COMBINE (Figure 6). COMBINE is a standalone application designed to track users' activities by connecting application nodes (app nodes) automatically.[9] After inserting a MMP network viewer, the double-click action adds a structure node. Users can look up known activities or generate fragments and perform a fragment search to identify a compound forming the MMP with the starting compound. The knowledge network formed is a concise and intuitive way of describing actions performed. Annotations can be added and aid users but do not require to follow because arrows linking app nodes indicate directionality and intention. The knowledge network saved by a user can be easily shared with other users, which are not possible with the web application.

**Large-Scale Compound-Target Relationship Analyses**

A connection between a compound and a target is typically assessed by examining the protocol of an assay the compound is active against. The connection is made if the assay protocol indicates that the activity reflects the binding of the compound to the protein target. A concentration cutoff is typically used to define what is "active", but the value is somewhat arbitrary and changes depending on the system and the type of the question one wants to answer. Rather than setting an arbitrary concentration cutoff to assess a compound-target connection, a compound-assay connection is first established instead, if the compound is tested against the assay regardless of the screening result. If the assay is connected to a protein target, a compound-target connection is inferred. Thus, a compound (C) – assay (A) – target (T) connection simply describes the compound testing event which involves screening C against A which is associated with T. To reduce the number of nodes in the network, C-A-T is shortened to C-T, and it represents a compound and an intended target. Connections were made this way for two reasons. The first reason was to not to introduce the activity cutoff concentration, essentially eliminating the need to define what is "active" and avoiding the removal of false negatives. The second reason was to include as many compounds as possible to examine diverse chemotypes introduced during the testing stage and to focus on the intention behind the selection or synthesis of compounds. The goal is to use the MMP network as an idea generating tool, a starting point to explore different chemotypes and their potential activities.

To reduce the number of colors used in the MMP network, nodes were colored according to their intended target classes (ITCs) rather than individual protein targets. Eight protein target classes, CYP450, GPCR, ion channel, kinase, nuclear, PDE, phosphatase, and protease, were colored in red, green, sky blue, blue, orange, violet, purple, and yellow, accordingly. If the number of ITCs of a compound is more than one, the node representing the compound is colored in black to indicate that it belongs to multiple classes. If a compound does not belong to one of eight ITCs, the node representing the compound is colored in gray. Finally, the node representing a ring fragment in Ring network is colored in dark green. Figures 2f-g show five different MMP networks, All, Center, GPCR, Kinase, and Ring networks, with ITCs colored as described.

Many small connected sets spread around bigger ones. This is the characteristic LGL layout pattern. Sets contain either a single ITC (single color except for black and gray) or mixture of many ITCs (multiple colors). Small connected sets with a single color tend to be originated from a scientific paper describing evaluation of biological activities of analogues synthesized using a small number of scaffolds. A lack of other ITCs indicates that compounds are most likely novel. The presence of small number of other ITCs is an interesting one because other potential activities can be inferred. Figure 7 shows a small connected set containing GPCR ITC compounds (adenosine receptor binders) except for one Kinase ITC compound (tyrosine-protein kinase SRC binder). The weak kinase activity associated with CHEMBL471913 indicates that other adenosine receptor binders might exhibit similar kinase activity.

A higher percentage of mixed ITCs can mean many things without examining activities of compounds. If they are clustered like ones found in Figure 8, it could mean that compounds like CHEMBL1823249 can be modified to modulate multiple targets. Modifying moieties A, B, and C of CHEMBL1823249 (Figure 8a) exhibits three different activities, protein-tyrosine phosphatase, hepatocyte growth factor receptor, and Cannabinoid CB2 receptor (Figure 8b).

The center of the All network is a highly connected set made up of about 44 % of the total number of compounds used to create various MMP networks. Such large connected network is possible when compounds forming MMPs contain a high profile/frequency chemotype, leading to many analogs. The presence of a small sized compound that can potentially be a substructure to many compounds can also lead to many MMPs. Because of the high number of connections present, laying out the network in visually unclutter way was not possible. To increase the resolution, compounds found only in the center set were extracted to create the Center network. One way to break up the Center network is to put a condition to not form a MMP with compounds involved in unusually high number of MMPs or with small sized compounds. Another way is to reduce the number of compounds by looking at one type of ITC. Or both approaches can be implemented to further reduce the possibility of forming highly connected networks. Figure 9 shows two reduced compound number networks created by considering GPCR and Kinase ITCs, two target classes of a high importance. The highly connected center sets still exist due to small sized compounds (Figure 9a), but they are more organized (edges formed are shorter). Many small sets containing mostly one color (green or blue) tend to include compounds coming from a single reference source as before. Ones found around the outer edge tend to be made up of compounds with complex ring systems or unusual differences like changing from S-S to Se-Se (Figure 9b). Such MMPs are infrequent and are reflected in the networks.

An interesting relationship one can analyze is to generate ring fragment to compound pairs. The MMP generating process produces fragments generated by breaking one, two, or three bonds (Figure 4). Ring fragments and compounds containing them can be used to generate Ring network shown in Figure 10. The resulting network can be used to quickly identify compounds sharing same set of ring fragments. A network with a low dark green node to other color node ratio represents a group containing simple ring frameworks or ring fragments with few variable positions. This is an example in which two types of nodes are used: compound and ring fragment. It can be extended to include multiple subtypes. For example, it is possible to create a hierarchy between two fragments in addition to a compound-fragment relationship. If a compound (C) contains four fragments (F1, F2, F3, and F4), one can extract pair relationships (C-F1, F1-F2, F2-F3, F2-F4, etc.) which can be used to create a hierarchy. Figure 11 illustrates how a hierarchy containing network can be formed using frameworks. A framework is simply one or more rings containing a linker between them (Figure 11a). The relationships between a compound and frameworks are listed in Figure 11b. The resulting hierarchy generated using a single compound is shown in Figure 11c. The process can be easily applied to a large number of compounds to examine different frameworks and their therapeutic importance.

**CONCLUSION**

Visualization is an important part of data analysis, and having a right visual representation can easily uncover hidden relationships that can be difficult to discover with other methods. To be effective, a visualization system needs to handle the big picture (to see how local relationships interact) and to quickly process a focused area with detailed information when needed. When the number of visual items increases, the ability to render the big picture decreases rapidly. One simple way to deal with this problem is to prerender the content, and the PRECUT process described herein shows how this can be achieved by applying the map display technology.[6,7,8] Tiles generated during the process capture a lot of information and can be delivered efficiently to users. The key to make this approach truly work, however, is to link detailed information in real time to complement the prerendered content. A solution implemented in a web application works but is cumbersome because linked information is presented in a separate web page, leading to many pages that cannot be tracked easily (Figure 5). A better solution is one implemented in COMBINE which connects different app nodes to track which activities were carried out (Figure 6). Information that can be linked are not limited to app nodes containing texts. App nodes containing other types of information such as chemical structures, images, and protein structures are possible (Figure 12). Generating pairs with different combinations of pair types and applying other layout algorithms are being actively investigated. Combining a content created using a layout algorithm and an infographic or linking multiple prerendered contents could be an interesting visual experience.

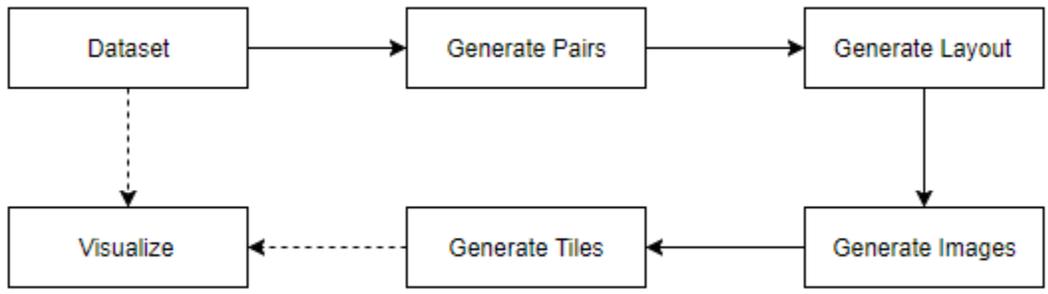

Figure 1. PRECUT process. Dashed arrows indicate that a visualizer uses data obtained from the original dataset and generated tiles to deliver a content.

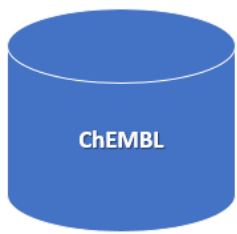
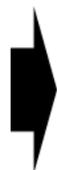
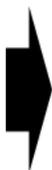
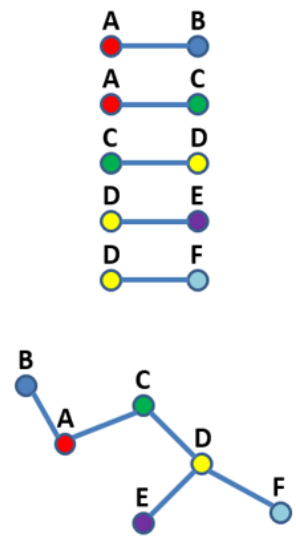

(a)            (b)            (c)

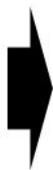
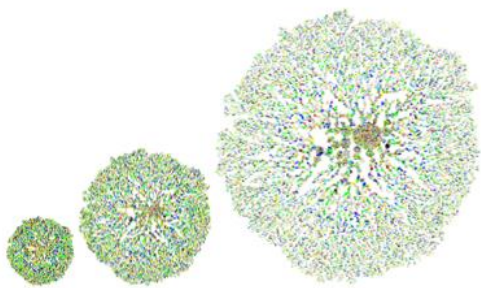
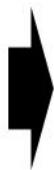
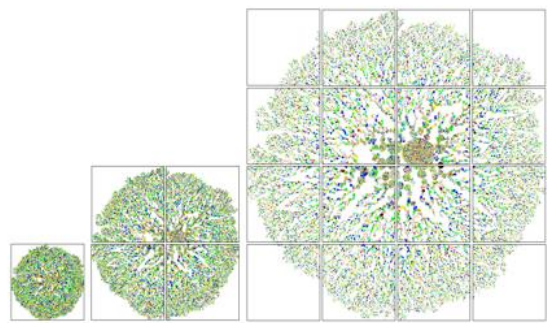

(d)            (e)

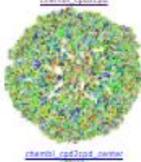

(f)

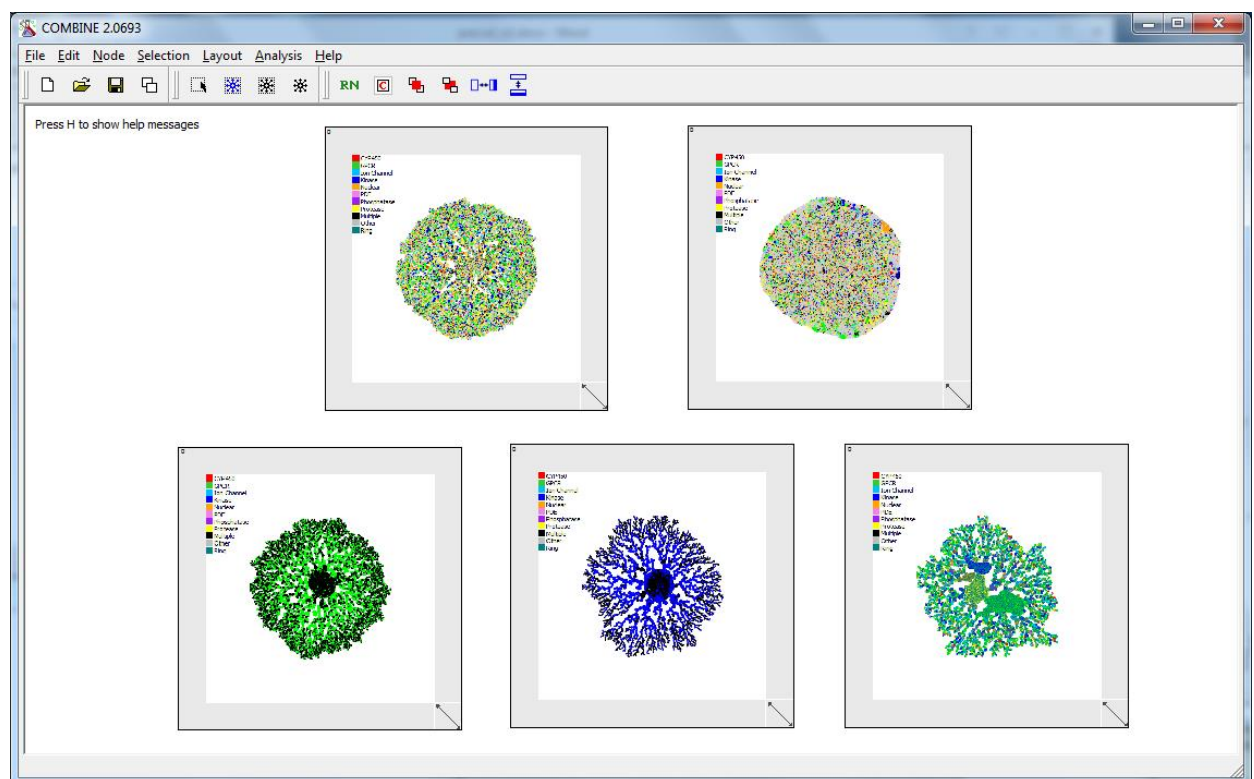

(g)

Figure 2. The PRECUT process applied to ChEMBL compounds. (a) 782,524 compounds were selected from the ChEMBL database. (b) 14,680,477 MMPs were generated after applying single, double, and triple cuts (Figure 3) and index them. (c) The LGL program is used to form graphs and calculate 2D coordinates of nodes. (d) 256x256, 512x512, and 1024x1024 pixel images were generated representing zoom levels 0, 1, and 2, respectively. (e) 1, 4, and 16 tiles were generated from three images in (d). (f) The MMP network viewer hosted at http://cheminformatic.com/mmpnet/. From the top, All network (782,524 compounds with 14,680,477 MMPs), Center network (341,052 compounds with 12,050,339 MMPs), GPCR network (253,024 compounds with 5,041,391 MMPs), Kinase network (61,405 compounds with 567,488 MMPs), and Ring network (260,711 compounds with 2,605,526 MMPs). (g) The MMP network viewer implemented in COMBINE with five networks described in (f).

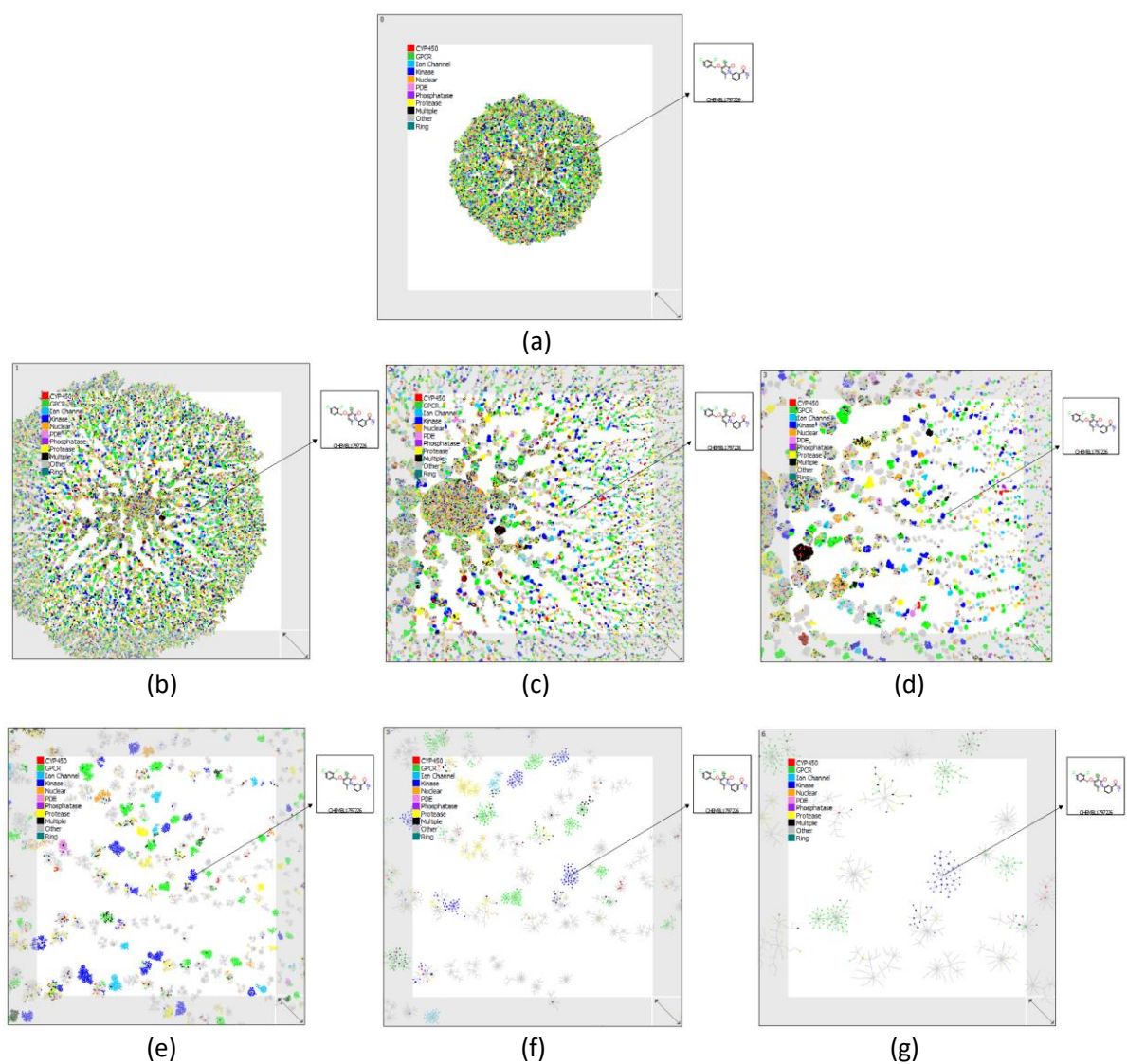

Figure 3. How tiles generated for different zoom levels are used by the MMP network viewer is illustrated. The top level (a) and zoomed levels 1-6 (b-g) are shown. The mouse scroll bar is used to control the zoom level.

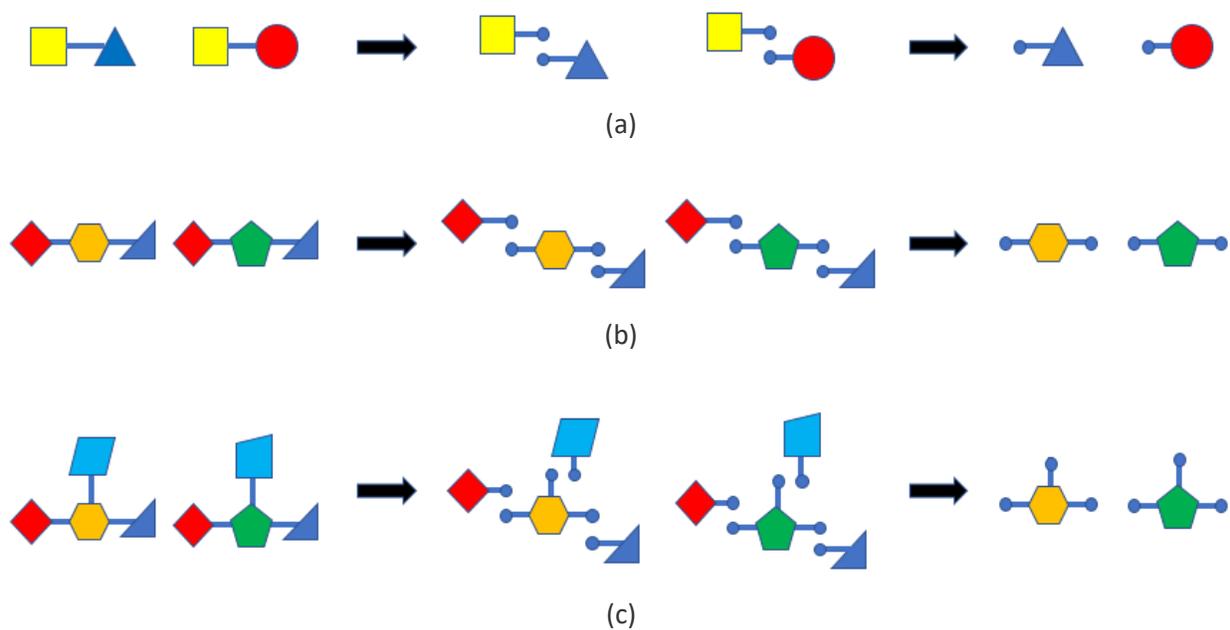

Figure 4. Single (a), double (b), and triple (c) cuts allowed during the MMP generating process. The "key" fragments are ones common to both, and the "value" fragments are ones that are different (shown after the second arrow).

(a)

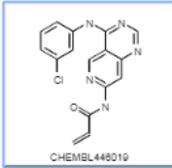

(b)

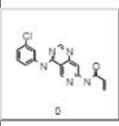

(c)

(d)

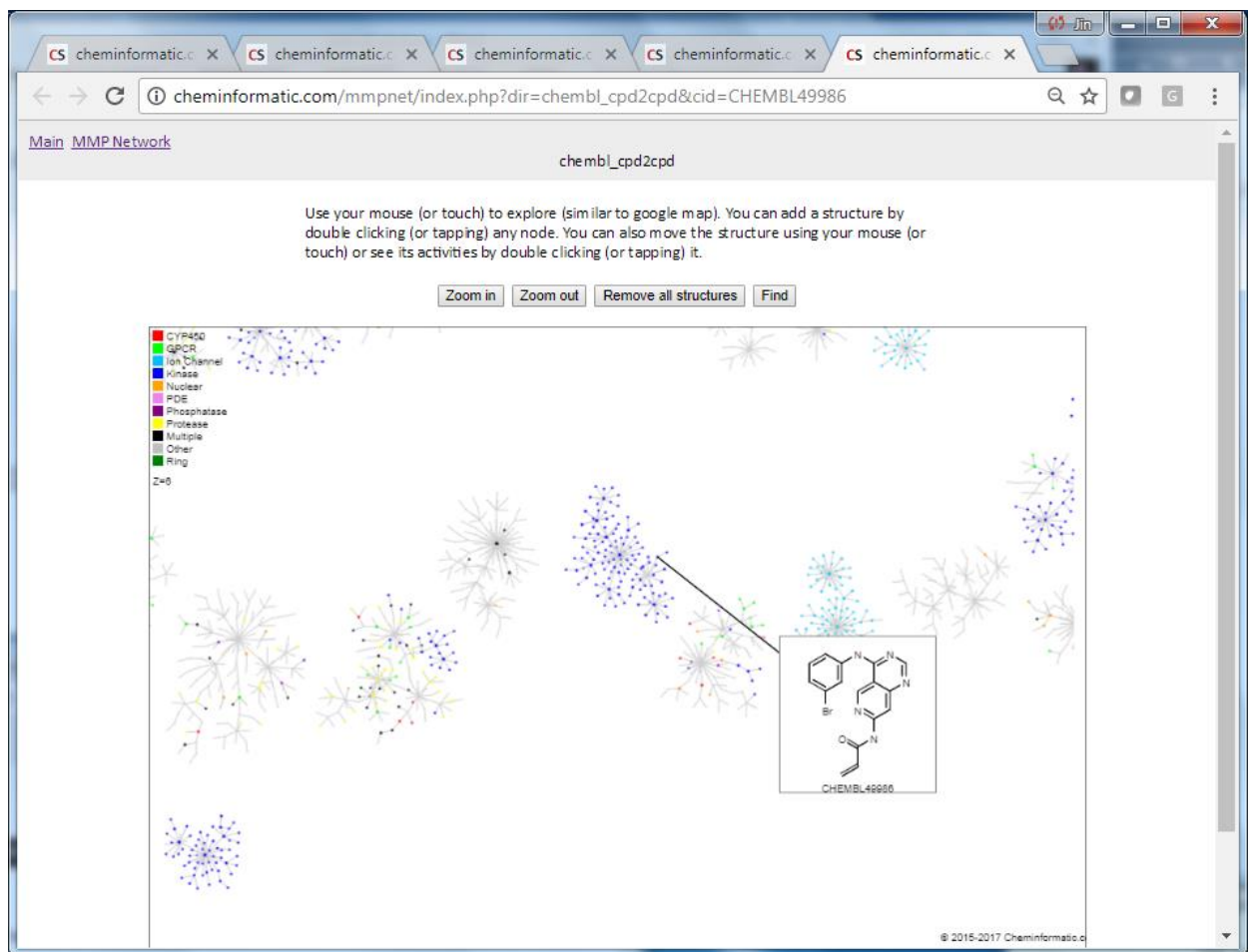
(e)

Figure 5. The MMP network viewer hosted at http://cheminformatic.com/mmpnet/. How to use the viewer is illustrated. (a) Double clicking a node generates the structure of selected ChEMBL compound. The structure is linked to the node by a line, and double clicking the structure creates a table containing the pharmacology data. (b) The pharmacology data is summarized in a table. Clicking the structure button on the top generates fragment buttons. (c) Clicking the button performs a fragment search. (d) ChEMBL compounds containing the fragment are listed. Two links are displayed on the top of each ChEMBL compounds. The first one points to the ChEMBL web site. The second one opens the MMP network viewer. (e) The MMP network viewer with the structure of selected compound.

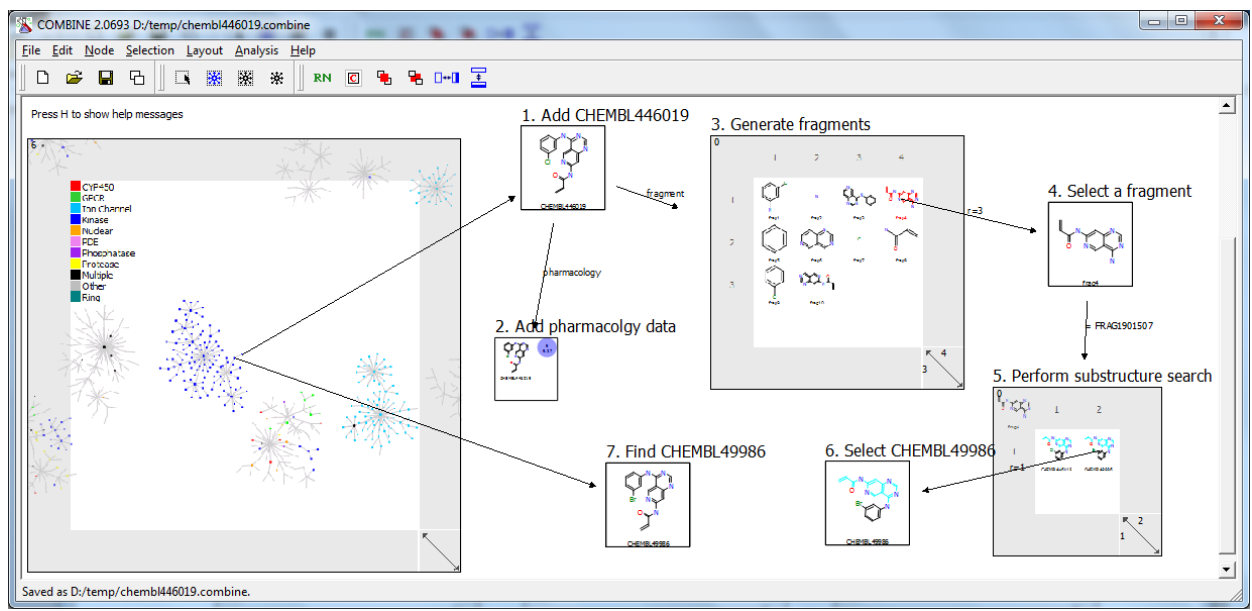

Figure 6. A knowledge network created using the MMP network viewer (left) and other app nodes. It shows how to find a MMP formed using CHEMBL446019 and CHEMBL49986.

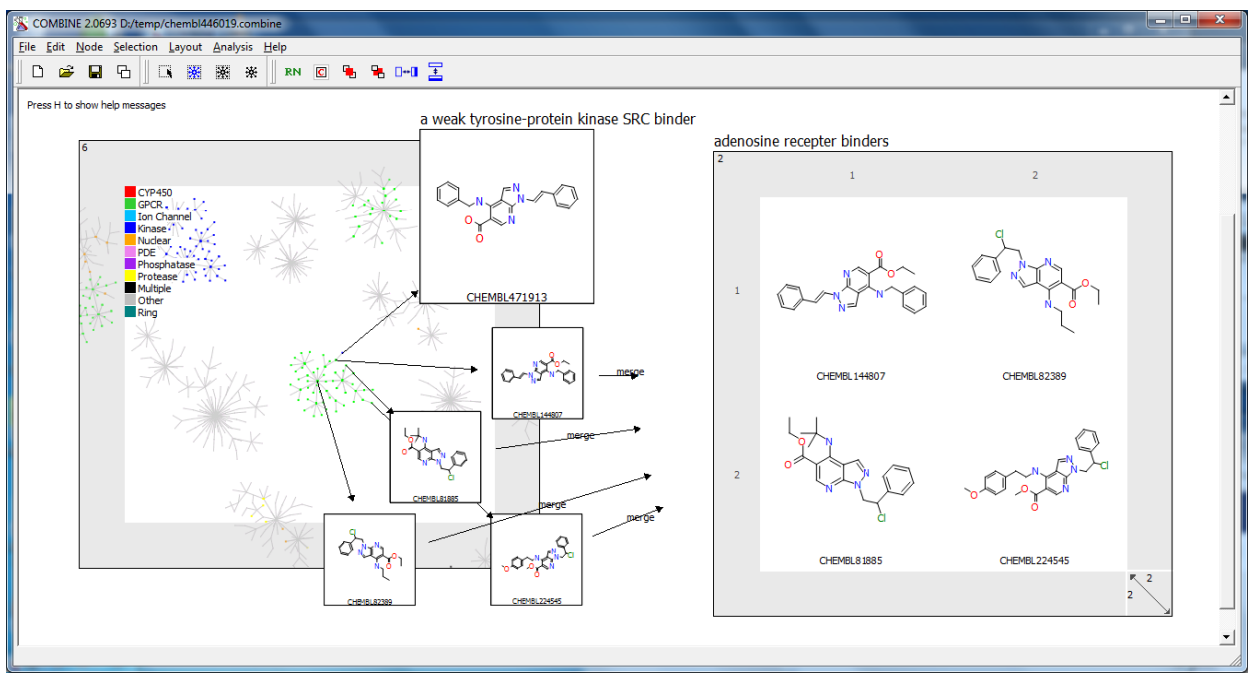

Figure 7. A connected network in All network containing adenosine receptor binders and a tyrosine-protein kinase SRC binder

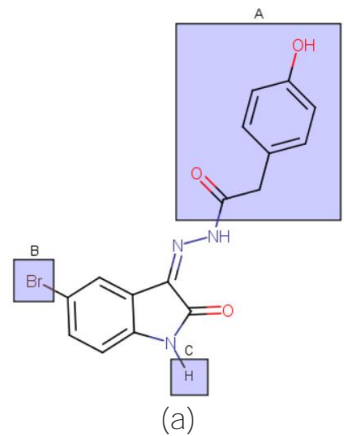

(a)

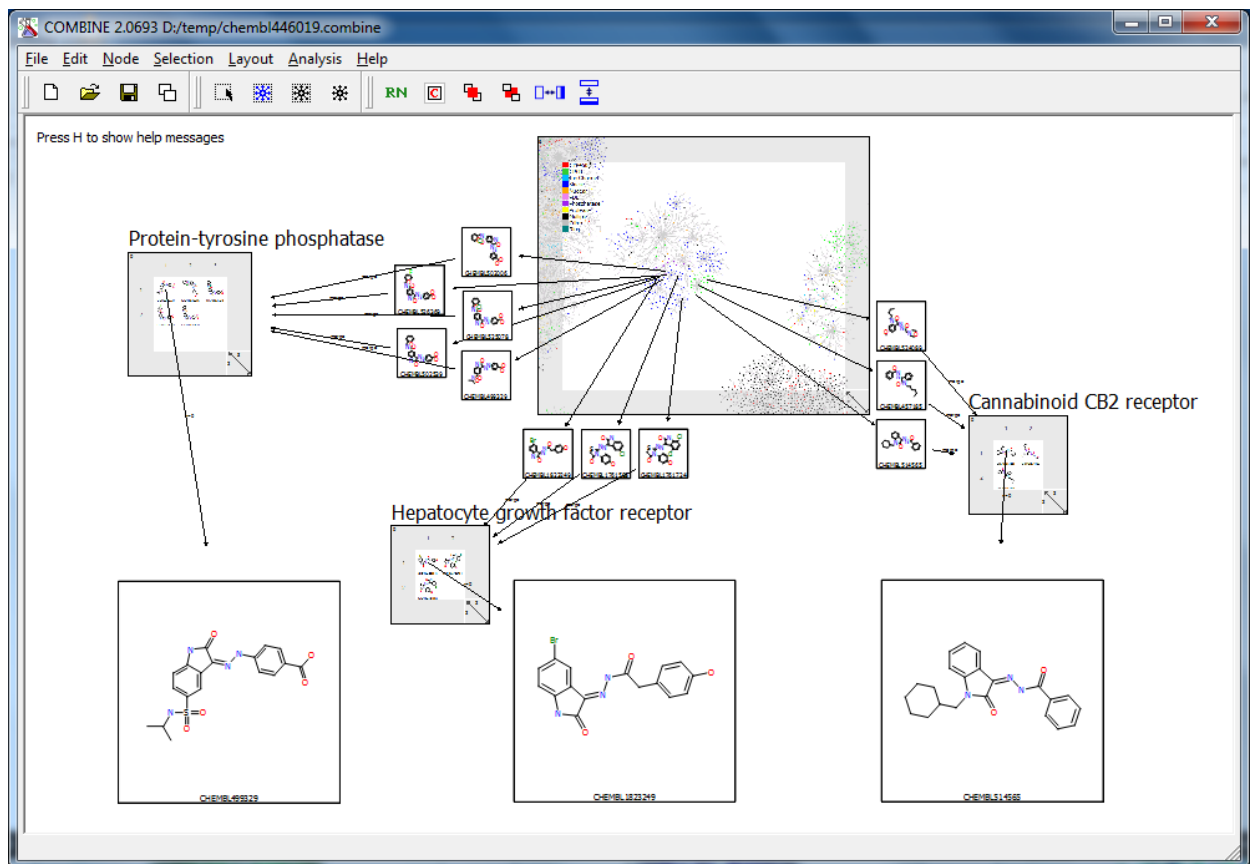

(b)

Figure 8. An example of how structure changes lead to different intended target classes found in All network. (a) The structure of CHEMBL1823249. Changing moieties, A, B, and C leads to different intended target classes. (b) A connected network containing three intended target classes, phosphatase, kinase, and GPCR.

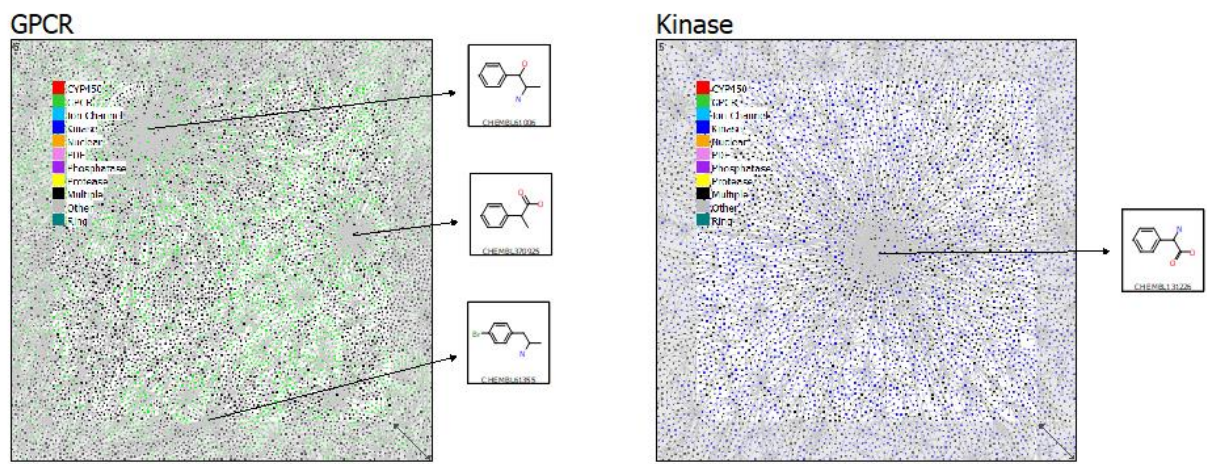

(a)

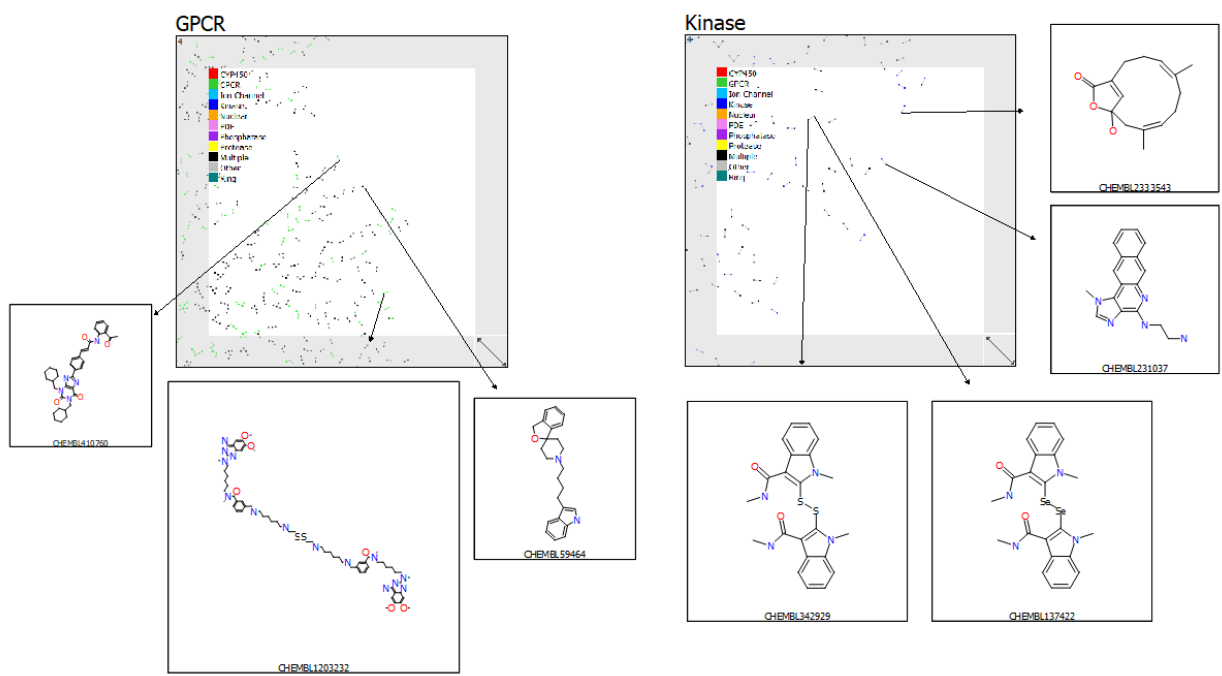

(b)

Figure 9. GPCR and Kinase networks. (a) Connected sets at the center. (b) Small connected sets near the outer edge.

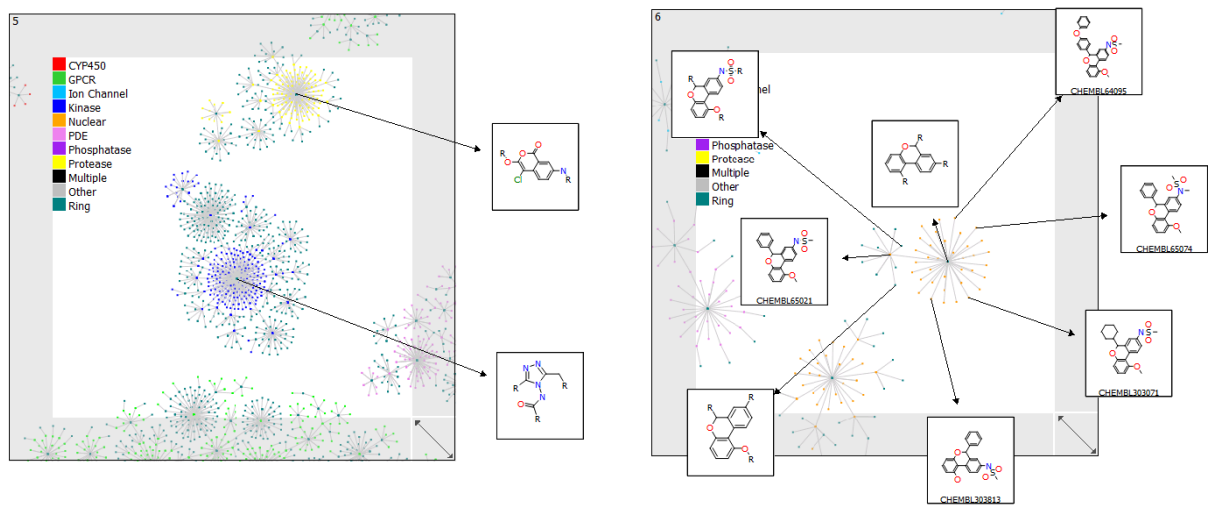

Figure 10. Ring fragments and compounds containing them.

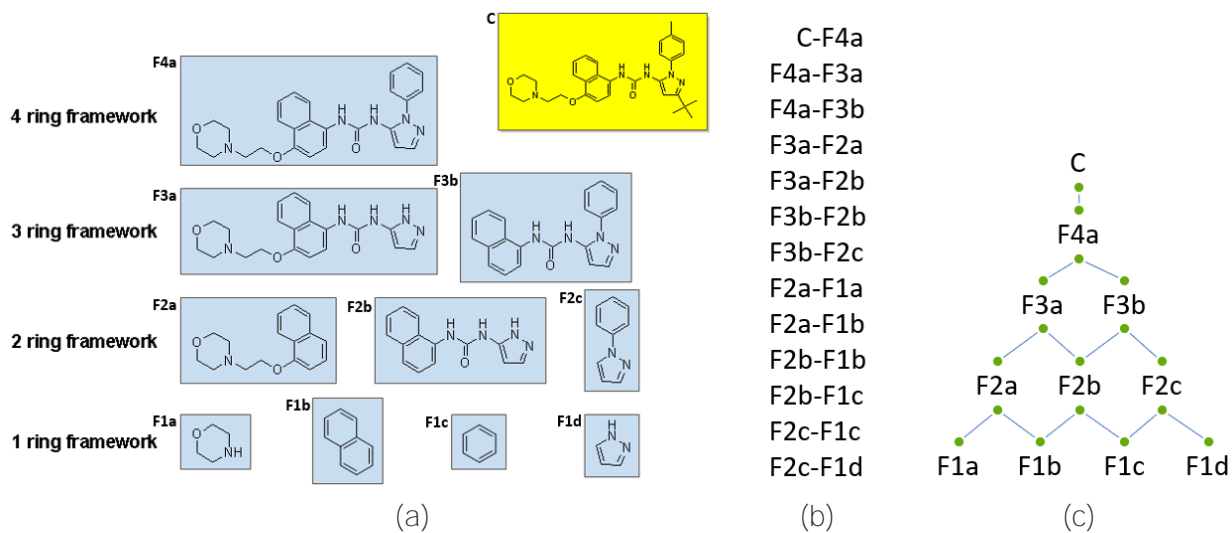

Figure 11. A framework hierarchy generated using p38 MAP kinase inhibitor BIRB 796.[17] (a) Different ring frameworks are generated from BIRB 796. (b) Pairs are formed from them. (c) A hierarchy is generated from pairs.

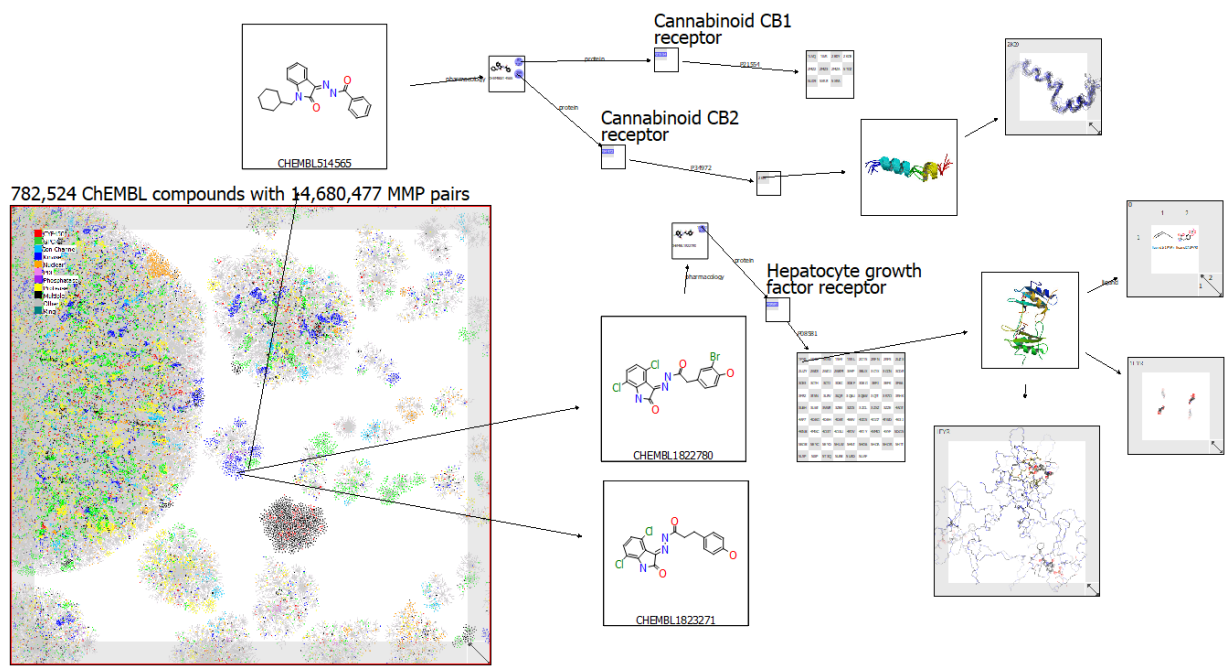

Figure 12. The MMP network viewer is used as a starting app node to generate this knowledge network. After displaying the structures of three ChEMBL compounds, the pharmacology data of two ChEMBL compounds were added.